# Online Thevenin Equivalent Parameter Estimation using Nonlinear and Linear Recursive Least Square Algorithm

Md. Umar Hashmi
Department of Energy Science & Engineering
IIT Bombay
Bombay, Maharashtra
umar.hashmi123@gmail.com

Rahul Choudhary
Department of Systems and Control
IIT Bombay
Bombay, Maharashtra
rahulchoudhary.2002@gmail.com

Jayesh G. Priolkar
Department of Electrical Engineering
Goa College of Engineering
Farmagudi, Ponda Goa
jayeshpriolkar@gmail.com

*Abstract*—This paper proposes method for detection, estimation of Thevenin equivalent parameters to describe power system behavior. Thevenin equivalent estimation is a challenge due to variation in system states caused by power flow in the network. Thevenin equivalent calculation based on changes in system with multiple sources integrated with grid, isolated distributed generator system is analysed and nonlinear least square fit estimation technique for algorithm is adopted. Linear least square fit is used with a linearized model. Performance evaluation of proposed method is carried out through mathematical model, nonlinear and linear least square fit based algorithm technique and simulation through MATLAB/SIMULINK package. Accurate grid and source side impedance estimation technique is applicable for distributed generation sources interfaced with grid to improve dynamic response, stability, reliability when subjected to faults or any other disturbances in network. Algorithm can accurately estimate Thevenin equivalent of multiple sources connected in parallel simultaneously with voltage and current phasor measurements at point of common coupling. Mathematical analysis and simulation results validate the effectiveness of proposed method.

## I. INTRODUCTION

With integration of large number of distributed generation sources (DG) into distribution network has raised concerns related to stability, power quality and reliability of the supply [1-3]. Distributed power generation sources depending upon their mode of operation affects line impedance of distributed electricity network. Performance of grid connected converter depends upon line impedance [4]. Islanding operation due to faults or grid disturbances due to voltage or frequency sag if not detected properly leads to poor power quality at point of common coupling (PCC).Online measurement due its numerous advantages is preferred to fulfill anti islanding requirement. The distortion in network caused due to electrical loads and distributed generation sources strongly depends upon grid impedance and amount of apparent power connected [5].To optimize the operation and performance of distributed generation sources in microgrid network, time and frequency dependent grid parameter estimation/ calculation is key [5].

In literature several methods are proposed for impedance estimation. Estimation based on control loop variations which include new values of inductance and resistance in control loop to improve performance of system is proposed in [6]. Grid impedance estimation based on use of extra devices is proposed in [7]. Injection of harmonics signal into the grid and use of mathematical models to obtain new impedance parameters are addressed in [8-10].Grid impedance using controlled excitation based on frequency characteristics of inductance -capacitance– inductance (LCL) filter resonance is reported in [3]. The various methods used to determine grid impedance characteristics offline by means of frequency response analysis is reported in [11-13].

The methods of equivalent impedance estimation can be broadly classified as active and passive type [14]. Passive methods utilize the existing disturbance present in power networks, for example detection of low order harmonic frequency impedance. In active methods in addition to regular operation, forced disturbance is injected into the grid or distributed generation network for parameter estimation. Regression analysis is used to extract parameters from measured data to define physical characteristics of system. Thevenin equivalent estimation based on concept of regression is also reported in literature. Recursive least square estimation technique based on varying system states to determine Thevenin equivalent is proposed in [15].

Detection and estimation of Thevenin equivalent impedance for different application is reported in literature. For example, power system fault detection, load matching for maximum power transfer, state of charge estimation for battery bank [16], simultaneous estimation of Thevenin equivalent of multiple sources, for load management by load shedding in power system network based on detection of undervoltage [17], voltage stability margin adjustment and analysis for prevention of voltage collapse by a real time voltage instability identification algorithm based on local phasor measurements [18].

The main aim of our work is to develop the algorithm for online Thevenin equivalent parameter estimation. It is active method which is available to estimate Thevenin equivalent voltage and current. To evaluate the performance of proposed method, mathematical model, nonlinear least square fit based algorithm technique and simulation through MATLAB/SIMULINK package are done.

Paper is organized as follows, Introduction of paper is given in section I, section II discusses in brief applications of online Thevenin equivalent estimation algorithm, section III presents parameter estimation for isolated power system, and section IV discusses the parameter estimation for grid connected system. Simulation results are discussed in section V. Section VI concludes the paper.

## II. APPLICATIONS OF THEVENIN EQUIVALENT ESTIMATION

Online Thevenin equivalent estimation algorithm finds application in adaptive control of electrical processes for performance improvement and preemptive corrective action can be taken for safe system operation. Some of the applications have been discussed in this section.

i) **Power system fault detection**: - The Thevenin equivalent of power system will indicate the change of estimated parameters due to fault in power system. The relationships between the real fault distance and the varying Thevenin equivalent impedance is presented in paper [19]. The proposed Thevenin equivalent estimation algorithm in our work uses balanced symmetrical components for correct estimation. Balanced ground faults, balanced short circuits and balanced impedance change can be identified using the presented algorithm.

ii) **Load matching for maximizing power transfer**: - Maximum power is transferred to a load resistor ($R_L$) when the value of the load resistor is selected to match the value of the Thevenin resistance ($R_{th}$) of the power source. Figure 1 demonstrates this for a value of $R_{th}$ of 10 ohms.

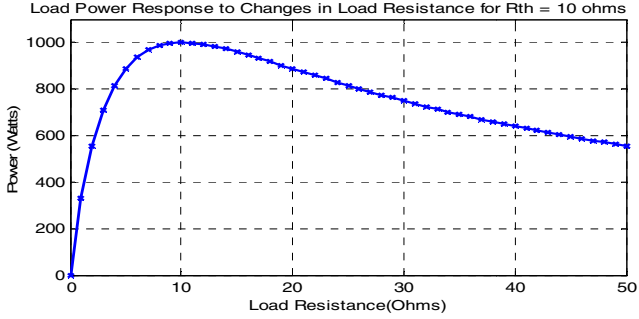

Figure 1: Maximum Power Transfer at $Z_{load}=Z_{source}$

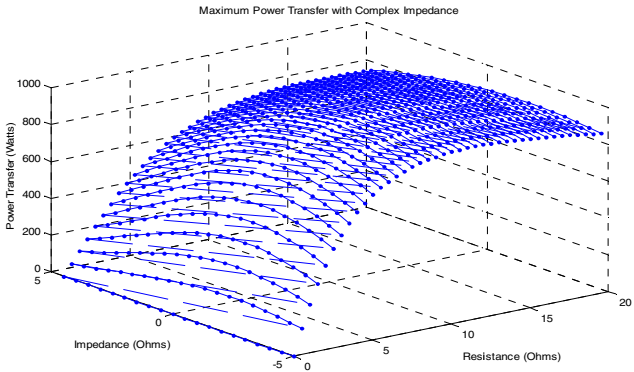

Figure 2: Maximum Power Transfer with complex impedance

$P = I^2 R$, $P_L = [V_{th}/(R_{th}+ R_L)]^2(R_L)$ (1)

One example is load matching is for wireless power transfer circuitry design using inductive coils. Four coil wireless power transfer simulation diagram along with the matching network is shown in Figure 3. Parameter values of the matching network are dependent on the Thevenin equivalent impedance of the four coil system.

Matching network shown in figure 3 is one of the many possible architectures of a matching network.

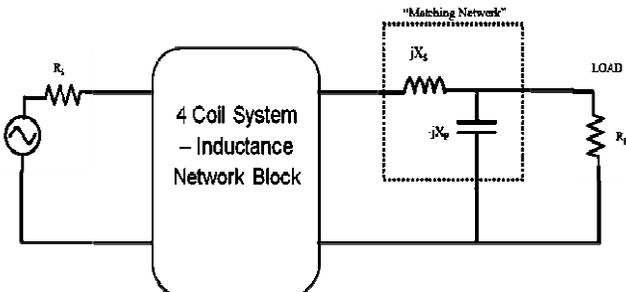

Figure 3: Matching network design for 4 coil wireless power transfer circuit

iii) **State of charge estimation of battery or battery bank**: A linear relationship is presented in [16] between the open circuit voltage (VOC) and SOC

$$SOC(\%) = \frac{v_{oc}(t) - b}{a}$$

The values of *a* and *b* are found experimentally [16]. Open circuit voltage is the Thevenin equivalent voltage of a battery under no load condition.

iv) **Voltage stability margin adjustment and analysis for prevention of voltage collapse**: Voltage instability is a major concern for power systems operation. A Real-Time voltage instability identification algorithm Based on Local Phasor Measurements is proposed in [18] which recognize instabilities in voltage. The power transferred to the bus reaches its voltage stability limit when that Thevenin impedance has the same magnitude as the load impedance at the bus [18].

The apparent power supplied is S, $Y = 1/Z_L$, than

$$\frac{dS}{dY} = \frac{E_{th}^2(1-Y^2 Z_{th}^2)}{(1+Z_{th}^2 Y^2 + 2Z_{th} Y \cos(\theta-\varphi))^{\wedge}2} \quad (2)$$

The condition for maximum load apparent power is dS/dY =0, hence the critical point of voltage instability is $Z_{th} = Z_L$. Hence maximum loading point can be accurately monitored online by calculating dS/dY. The value of dS/dY close to zero indicates proximity to voltage collapse point [20].

The proposed algorithm assumes no simplification in system for estimating Thevenin equivalent parameters.

### III. PARAMETER ESTIMATION FOR ISOLATED SYSTEM

**Description of the Proposed Algorithm:**
Consider a voltage source feeding a load which has an impedance of itself, line and filter impedance before it's actually connected to the load. The Thevenin equivalent of the circuit is shown adjacent to it in the figure below.

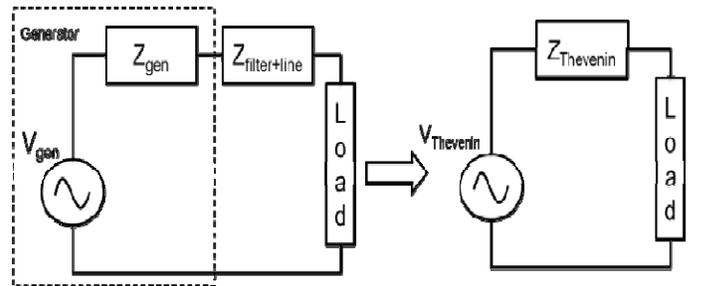

Figure 4: Isolated system with its Thevenin equivalent

Using Kirchhoff's Voltage Law we can write the voltage loop equation as:

$V_{\text{Thevenin}}(\cos \theta + j \sin \theta) = V_{PCC}(\cos \emptyset + j \sin \emptyset) + I_{PCC}(\cos \varphi + j \sin \varphi) \times (R_{\text{Thevenin}} + jX_{\text{Thevenin}})$ (3)

We can separate the real and imaginary component of the equation as,

$V_{\text{Thevenin}} \cos \theta = V_{PCC} \cos \emptyset + I_{PCC} (\cos \varphi \times R_{\text{Thevenin}} - \sin \varphi \times X_{\text{Thevenin}})$ (4)

$V_{\text{Thevenin}} \sin \theta = V_{PCC} \sin \emptyset + I_{PCC}(\cos \varphi \times X_{\text{Thevenin}} + R_{\text{Thevenin}} \times \sin \varphi)$ (5)

The algorithm will be able to estimate the Thevenin equivalent voltage and current. The point of common coupling is the point at which the load is connected. The known and unknown parameters are shown below in table 1:

Table 1: State of Parameters

| Known Parameters | Unknown Parameters |
|---|---|
| Voltage Magnitude at PCC ($V_{PCC}$) | Thevenin Voltage ($V_{Thevenin}$) ($x_1$) |
| Voltage Angle at PCC ( ) | Thevenin Voltage Angle ($\theta$) ($x_2$) |
| Current Magnitude at PCC ($I_{PCC}$) | Thevenin Resistance ($R_{Thevenin}$) ($x_3$) |
| Current Angles at PCC ($\varphi$) | Thevenin Inductance ($X_{Thevenin}$) ($x_4$) |

$$V_{PCC} \cos \emptyset = V_{Thevenin} \cos \theta - I_{PCC}(\cos \varphi \times R_{Thevenin} - \sin \varphi \times X_{Thevenin}) \quad (6)$$

$$V_{PCC} \sin \emptyset = V_{Thevenin} \sin \theta - I_{PCC}(\cos \varphi \times X_{Thevenin} + R_{Thevenin} \times \sin \varphi) \quad (7)$$

Equation (6) and (7) are non-linear in nature.

$$\begin{bmatrix} y_1 \\ y_2 \end{bmatrix} = \begin{bmatrix} x_1 \cos(x_2) - ax_3 + bx_4 \\ x_1 \sin(x_2) - bx_3 - ax_4 \end{bmatrix} \quad (8)$$

Where, $y_1 = V_{PCC} \cos \emptyset$, $y_2 = V_{PCC} \sin \emptyset$, $a = I_{PCC} \cos \varphi$, $b = I_{PCC} \sin \varphi$ are known parameters and $x_1 = V_{Thevenin}$, $x_2 = \theta$, $x_3 = R_{Thevenin}$, $x_4 = X_{Thevenin}$ are unknown parameters.

$$Y = f(X, a, b) \quad (9)$$

Where $Y = \begin{bmatrix} y_1 \\ y_2 \end{bmatrix}$ and $X = \begin{bmatrix} x_1 \\ x_2 \\ x_3 \\ x_4 \end{bmatrix}$

$$\beta = \min_X \sum_{k=1}^{2 \text{ or greater}} [Y - f(X)]_k^2 \quad (10)$$

It is intended to solve these equations to obtain the values X which satisfy the system of equations described in equation 10. Initial guess values are random numbers. Equation 10 has to be minimized for optimal values of unknown parameters. Ideally β has to be zero but since the algorithm recursively optimizes the unknown parameters, there is a residual error which has to be maintained as low as possible for accurate results. Residual is defined by equation (11).

$$Residual = Y - f(X_{optimal}) \quad (11)$$

Nonlinear Recursive algorithm needs two Phasor measurements of voltage and current at the point of common coupling. The figure below shows the circuit diagram for two different loading scenarios. Phasor measurements are done at steady state.

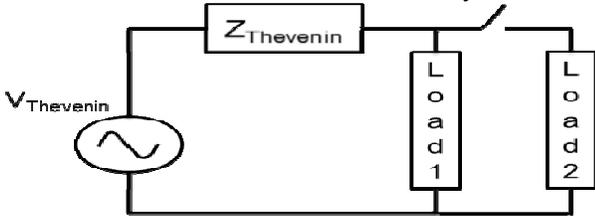

Figure 5: Circuit description of two data set collection

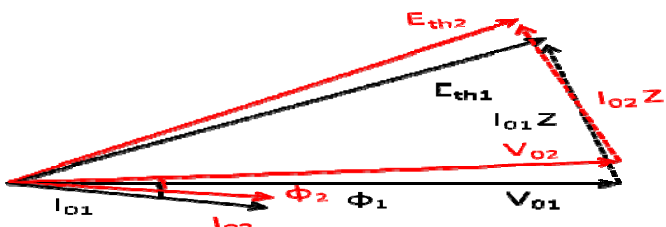

Figure 6: Phasor diagram of at least two different measurements is required for operation of estimation algorithm.

Estimation of Thevenin equivalent uses local measurements of at least two different voltage and current vectors (magnitude and phase) pairs measured at different time with different values associated with same reference Thevenin equivalent parameters.

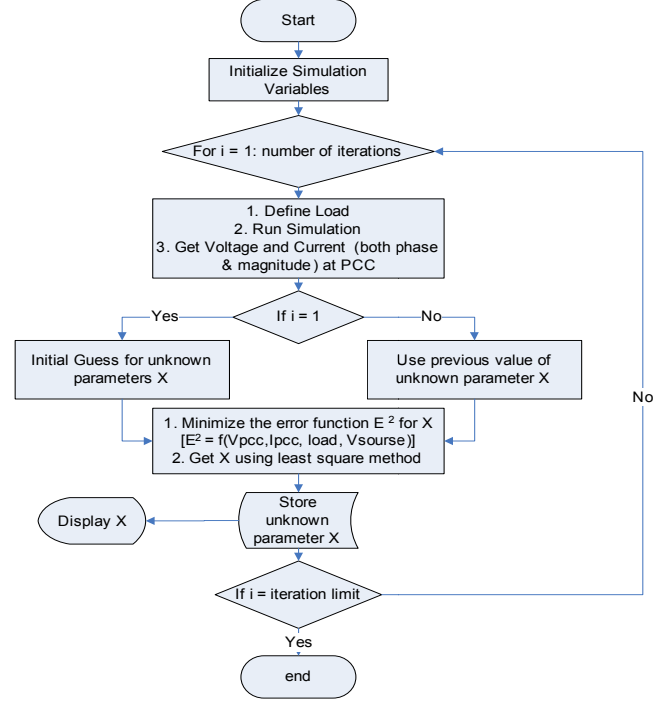

Figure 7: Flow chart for the proposed algorithm

**Parameters Estimation using linear regression**

Non-linearity of the Equation (8) can be taken care by eliminating the trigonometric variables.

$$\begin{bmatrix} y_1 \\ y_2 \end{bmatrix} = \begin{bmatrix} x_1 - ax_3 + bx_4 \\ x_2 - bx_3 - ax_4 \end{bmatrix} \quad (12)$$

$$\begin{bmatrix} y_1 \\ y_2 \end{bmatrix} = \begin{bmatrix} 1 & 0 & -a & b \\ 0 & 1 & -b & -a \end{bmatrix} \times \begin{bmatrix} x_1 \\ x_2 \\ x_3 \\ x_4 \end{bmatrix} \quad (13)$$

Where, $y_1 = V_{PCC} \cos \emptyset$, $y_2 = V_{PCC} \sin \emptyset$, $a = I_{PCC} \cos \varphi$, $b = I_{PCC} \sin \varphi$ are known parameters and $x_1 = V_{Thevenin} \cos \theta$, $x_2 = V_{Thevenin} \sin \theta$, $x_3 = R_{Thevenin}$, $x_4 = X_{Thevenin}$ are unknown parameters.

Additional variable can be calculated as

$$V_{Thevenin} = \sqrt{(x_1^2 + x_2^2)} \quad (14)$$

$$\theta = \tan^{-1} \frac{x_2}{x_1} \quad (15)$$

For n different loads, Equation (13) can be written in following structures (augmentation of matrix),

$$\begin{bmatrix} Y_1 \\ Y_2 \\ \vdots \\ Y_n \end{bmatrix} = \begin{bmatrix} A_1 \\ A_2 \\ \vdots \\ A_n \end{bmatrix} X \quad (16)$$

Where dimension of the matrices are as, $Y_{2n \times 1} = A_{2n \times 4} X_{4 \times 1}$

For given initial estimate of $X = \hat{X}$,

$$\hat{Y} = A\hat{X} \quad (17)$$

Estimation error

$$E = Y - \hat{Y} \Rightarrow Y - A\hat{X} \quad (18)$$

$$\beta = E^2 = EE^T = [Y - A\hat{X}][Y - A\hat{X}]^T \quad (19)$$

Error between actual and estimate will be minimum for

$$\frac{\partial \beta}{\partial \hat{X}} \Rightarrow \frac{\partial([Y-A\hat{X}][Y-A\hat{X}]^T)}{\partial \hat{X}} = 0 \quad (20)$$

$$A^T A \hat{X} - A^T Y = 0 \quad (21)$$

$$\hat{X} = (A^T A^{-1}) A^T Y \quad (22)$$

Vector $\hat{X}$ provides good estimate of unknown parameters.

## IV. PARAMETER ESTIMATION FOR GRID CONNECTED SYSTEM

Consider DG source connected to grid as shown in figure 8. The total impedance of network includes line, filter and load impedance.

**Thevenin Estimation Results:**

Positive sequence voltage and current angle and magnitude are measured. It is assumed that the input voltage magnitude and angle and the Thevenin impedance as seen from the load side is constant while the algorithm is in operation. As the estimation gives an output which is within ±1% error with just one load change; and hence it does not take more than a few seconds for accurate prediction.

The positive sequence extraction is done to eliminate the unbalanced components in voltage and current. The impact of negative sequence is in terms of losses or heating caused by unbalances, for this case it's not that relevant.

The results of the estimation with multiple numbers of iterations are tabulated below table.

Table 2: Estimation of Parameters for Isolated system

| No. of Phasor Measurements | Generator Voltage | Voltage angle | Resistance | Inductance |
|---|---|---|---|---|
| Actual | 70.7107 | | 1 | 0.377 |
| 2 Phasor | 70.642 | 1.98E-05 | 0.98442 | 0.3747 |
| Error % | 0.09706 | | 1.55733 | 0.5993 |
| 3 Phasor | 70.642 | 2.98E-05 | 0.984399 | 0.374741 |
| Error % | 0.09715 | | 1.5601 | 0.5992 |
| 5 Phasor | 70.6254 | 1.11E-05 | 0.97855 | 0.373026 |
| Error % | 0.12057 | | 2.14427 | 1.05397 |
| 10 Phasor | 70.576 | 2.76E-05 | 0.9609 | 0.3691 |
| Error % | 0.1905 | | 3.91 | 2.0955 |
| 100 Phasor | 70.5762 | 1.34E-06 | 0.961 | 0.3691 |
| Error % | 0.19015 | | 3.89949 | 2.09503 |

The estimated values tabulated in Table 2 shows accurate prediction with just two Phasor measurements of known parameters at point of common coupling.

Superposition theorem implementation of proposed algorithm:

The algorithm can be used for multiple sources parameter estimation simultaneously. The information needed from the system is that we should know the currents coming out of each of the sources and the common voltage vector at point of common coupling.

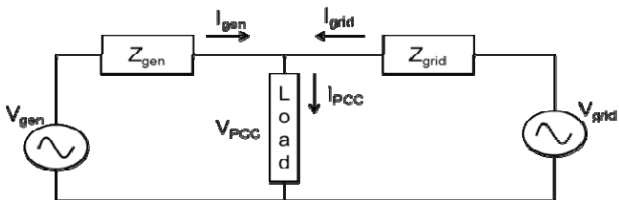

Figure 8: Generator and grid connected in parallel to feed a load

For parameter estimation for a system with multiple sources:
- Voltage magnitude and angle at PCC is needed.
- Current from individual sources ($I_{gen}$ and $I_{grid}$ in the figure above) is needed

Equivalent Thevenin voltage (magnitude and angle) and impedance can be estimated. The results of estimation are listed in Table 4 and 6. Generator Side Estimation known and unknown parameters are listed in table 3. A hypothetical scenario is simulated in which voltage levels of two sources connected in parallel are drastically different. The generator side voltage magnitude is set at 70.7 V and the grid is assumed at 49.5 V. The algorithm accurately estimates the voltage level difference.

Table 3: State of Parameters on Generator side

| Known Parameters | Unknown Parameters |
|---|---|
| Voltage Magnitude at PCC | Generator Voltage |
| Voltage Angle at PCC | Generator Voltage Angle |
| Current Magnitude of generator | Generator resistance |
| Current Angles | Generator Inductance |

Parameter Estimated with their errors percentages for generator side parameter estimation is tabulated in Table 4.

Table 4: Estimation of Parameters with error on Generator side

| No. of Phasor Measurements | Generator Voltage | Voltage angle | Resistance | Inductance |
|---|---|---|---|---|
| Actual | 70.7107 | | 1 | 0.377 |
| 2 Phasor | 70.70932 | -6.285 | 0.999933 | 0.376977 |
| Error % | 0.00195 | | 0.0067 | 0.0061 |
| 3 Phasor | 70.70948 | -6.279 | 0.999942 | 0.376974 |
| Error % | 0.00172 | | 0.00577 | 0.007004 |
| 5 Phasor | 70.71005 | -6.290 | 0.999971 | 0.376978 |
| Error % | 0.00092 | | 0.00289 | 0.0059 |

Grid Side Estimation known and unknown parameters are listed in table 5.

Table 5: State of Parameters on grid side

| Known Parameters | Unknown Parameters |
|---|---|
| Voltage Magnitude at PCC | Grid Voltage |
| Voltage Angle at PCC | Grid Voltage Angle |
| Current Magnitude from grid | Grid Resistance |
| Current Angles | Grid Inductance |

Parameter Estimated with their errors:

Table 6: Estimation of parameters with errors on grid side

| No. of Phasor Measurements | Grid Voltage | Voltage angle | Grid Resistance | Grid Inductance |
|---|---|---|---|---|
| Actual | 49.4975 | | 0.5 | 0.0377 |
| 2 Phasor | 49.4973 | -0.0019 | 0.5 | 0.0377 |
| Error % | 0.0004 | | 0 | 0 |
| 3 Phasor | 49.49726 | -6.29 | 0.500017 | 0.037699 |
| Error % | 0.000483 | | -0.00334 | 0.00285 |
| 5 Phasor | 49.49754 | -6.29 | 0.499 | 0.03769 |
| Error % | -8.26E-05 | | 0.00112 | 0.00157 |

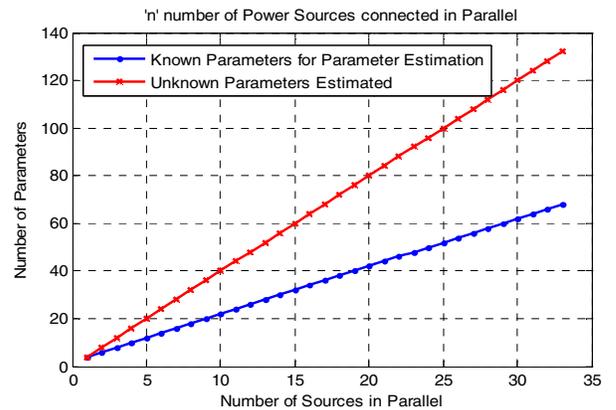

Figure 9: Known & Unknown parameters for 'n' sources connected in parallel

The results of estimation indicate accurate prediction of both generator and grid side parameters simultaneously. This algorithm can be extrapolated to 'n' number of parallel sources feeding a load. The algorithm uses the shared current from the source and voltage at point of common coupling for estimating Thevenin equivalent parameters. Figure 9 shows the number of estimated parameters and number of known parameters used for estimating the unknown values using the proposed algorithm. For very large number of sources connected in parallel, the algorithm proposed estimates twice the number of unknown parameters. This aspect of the proposed algorithm can help in controller design and tuning for multiple sources i.e. inverters and/or UPS to be connected in parallel. Virtual impedance design requires equivalent Thevenin impedance seen at from load side for proper compensation of parameter mismatches (Line and/or filter parameters). Simultaneous estimation of multiple sources Thevenin's equivalent will help in proper selection of virtual impedance for minimizing circulating current flowing among power sources.

Table 7: Parameter Estimation with linear regression

| No. of Phasor Measurements | Generator Voltage | Voltage angle | Resistance | Inductance |
|---|---|---|---|---|
| Actual | 70.711 | | 1.000 | 0.377 |
| 1 Phasor | 15.395 | 39.479 | -8.829 | -3.351 |
| Error % | 78.228 | | 982.878 | 988.831 |
| 2 Phasor | 47.984 | -30.161 | -1.342 | -6.005 |
| Error % | 32.140 | | 234.184 | 1692.739 |
| 3 Phasor | 70.602 | -0.151 | 1.005 | 0.390 |
| Error % | 0.154 | | -0.483 | -3.496 |
| 4 Phasor | 70.651 | -0.112 | 1.006 | 0.392 |
| Error % | 0.084 | | -0.557 | -4.019 |
| 5 Phasor | 70.599 | -0.143 | 0.999 | 0.374 |
| Error % | 0.158 | | 0.091 | 0.665 |
| 6 Phasor | 70.583 | -0.151 | 0.996 | 0.367 |
| Error % | 0.181 | | 0.374 | 2.730 |
| 10 Phasor | 70.636 | -0.110 | 0.998 | 0.371 |
| Error % | 0.105 | | 0.219 | 1.603 |

Linear recursive least square method derived for thevenin equivalent estimation requires at least three phasor measurements for accurate prediction of unknown parameters. The result for linear recursive least square estimation is listed in table 7. The results with 1$^{st}$ and 2$^{nd}$ phasor measurements are not reliable, however the results converge with 3 phasor measurements and own wards.

Compared to linear recursive fit, nonlinear recursive fit is better as the error percentages are much lower as per results tabulated in table 4, 6 and 7.

Nonlinear recursive method uses exact equation, therefore computationally exhaustive, for estimation and hence the overall accuracy of estimation is higher. Furthermore nonlinear recursion estimation converges with 2 phasor measurement data but linear method takes at least 3 phasor measurements for convergence.

The results generated using proposed algorithms has an upper bound of 5000 as maximum function evaluations and maximum number of iterations as 8000.

**Power System Grid Impedance Change Detection**

Power system undergoes various kinds of faults which lead to overall impedance change of the system. Simulations are conducted in MATLAB Simulink for isolated system to verify the capability of proposed algorithm to detect impedance change in power system. Single source simulations with impedance change from 1+0.377j to 2+0.755j at 5 second in a 9.6 second simulation is conducted. The algorithm is capable to detect the impedance change very accurately. As the source voltage and phase remains unchanged, the estimation remains fairly stable at its initial values.

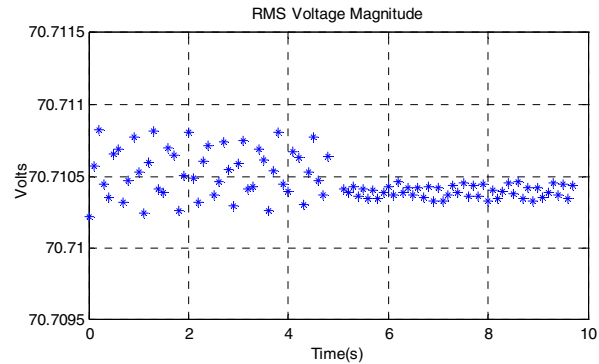
Figure 10: Voltage magnitude estimation

The voltage magnitude from source side is unchanged and only line impedance is changed. Therefore the Thevenin voltage magnitude and phase estimated value remains fairly stable.

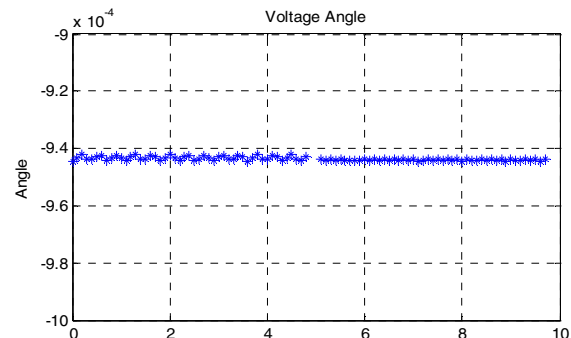
Figure 11: Voltage Phase estimation

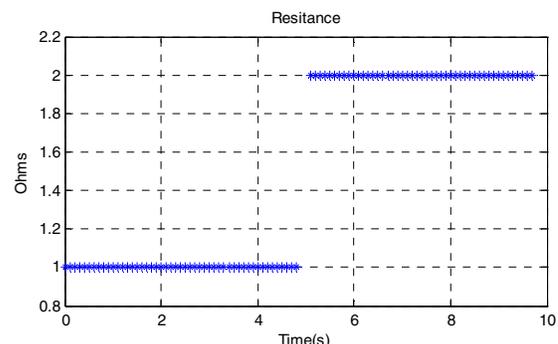
Figure 12: Power source Resistance estimation

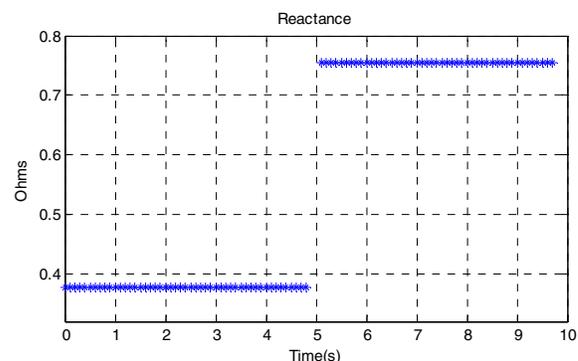
Figure 13: Power System reactance

As the resistance and reactance values are doubled at 5sec, the estimated values changes level around 5 sec. The error % for the estimated quantities is well below 0.5% under steady state. The positive sequence voltage and current angle and magnitude are measured. It is assumed that the input voltage magnitude and angle and the Thevenin impedance as seen from the load side is constant while the algorithm is in operation. The positive sequence extraction is done to eliminate the unbalanced components in voltage and current. Proposed algorithm can be used for detection of balanced power system faults using the proposed parameter estimation algorithm using linearized and nonlinear recursive least square algorithms evaluated through simulation results shown from Figure 10 to 13.

## VI. CONCLUSION

Online Thevenin equivalent parameter estimation using nonlinear and linear recursive least square algorithm are proposed and evaluated in this paper. Simultaneous parameter estimation of 'n' parallel source with known parameters at point of common coupling i.e. current and voltage phasor are evaluated and analysed. The proposed algorithms require at least two phasor measurements for accurate prediction. The robustness of the algorithms has also been tested and verified. Fault in power system leads to impedance change. The Thevenin equivalent estimation proposed in the paper can help to determine power system behavior and adopt the corrective actions to improve response of system network and stability. Simulation results validating the performance of the proposed algorithms have also been verified for power system impedance change detection. From the analysis carried out it is found that nonlinear recursion is comparatively better than linear recursion in terms of error percentage of the estimation. However linear method is computationally less exhaustive, more viable for online controller implementation.